\documentclass[pra,10pt,amsmath,twocolumn,floatfix,showpacs]{revtex4}
\usepackage{amsmath}
\usepackage{amsfonts}
\usepackage{graphicx}

\newcommand{\op}[1]{\operatorname{#1}}

\newcommand{\diff}[2]{\frac{d #1}{d #2}}

\newcommand{\intall}{\int_{-\infty}^{\infty}}
\newcommand{\ket}[1]{|#1\rangle}

\newcommand{\bra}[1]{\langle#1|}

\newcommand{\rect}{\operatorname{rect}}

\newcommand{\abs}[1]{\left|#1\right|}
\newcommand{\bk}[1]{\left(#1\right)}
\newcommand{\Bk}[1]{\left[#1\right]}
\newcommand{\BK}[1]{\left\{#1\right\}}

\newcommand{\trace}{\operatorname{tr}}

\begin{document}
%\twocolumn[
\title{Ziv-Zakai Error Bounds for Quantum Parameter Estimation}

\author{Mankei Tsang}

\email{eletmk@nus.edu.sg}
\affiliation{Department of Electrical and Computer Engineering,
  National University of Singapore, 4 Engineering Drive 3, Singapore
  117583}

\affiliation{Department of Physics, National University of Singapore,
  2 Science Drive 3, Singapore 117551}

%\author{Lorenzo Maccone}
%\affiliation{
%Center for Quantum Information and Control,
%University of New Mexico, MSC07--4220, Albuquerque, New Mexico
%87131-0001, USA}

%\author{Carlton M.\ Caves}

%\affiliation{
%Center for Quantum Information and Control,
%Department of Physics and Astronomy,
%University of New Mexico, Albuquerque, New Mexico
%87131, USA}

%\author{Seth Lloyd}

%\affiliation{Research Laboratory of Electronics,
%Massachusetts Institute of Technology, Cambridge, Massachusetts
%02139, USA}

%\affiliation{Department of Mechanical Engineering,
%Massachusetts Institute of Technology, Cambridge, Massachusetts
%02139, USA}

%\author{Jeffrey H.\ Shapiro}

%\affiliation{Research Laboratory of Electronics, Massachusetts
%  Institute of Technology, Cambridge, Massachusetts 02139, USA}

\date{\today}

\begin{abstract}
%  In classical statistics, the Ziv-Zakai bounds give lower limits to
%  the mean-square error in a parameter estimation problem by relating
%  it to the error probability in a binary hypothesis testing problem.
%  Compared with the more well known Cram\'er-Rao bounds, the Ziv-Zakai
%  bounds are often much tighter to the achievable error when the
%  likelihood function is highly non-Gaussian and the number of trials
%  is limited.

  I propose quantum versions of the Ziv-Zakai bounds as alternatives
  to the widely used quantum Cram\'er-Rao bounds for quantum parameter
  estimation.  From a simple form of the proposed bounds, I derive
  both a ``Heisenberg'' error limit that scales with the average
  energy and a limit similar to the quantum Cram\'er-Rao bound that
  scales with the energy variance. These results are further
  illustrated by applying the bound to a few examples of optical phase
  estimation, which show that a quantum Ziv-Zakai bound can be much
  higher and thus tighter than a quantum Cram\'er-Rao bound for states
  with highly non-Gaussian photon-number statistics in certain regimes
  and also stay close to the latter where the latter is expected to be
  tight.
\end{abstract}
\pacs{03.65.Ta, 42.50.St}

\maketitle
In statistics, one often has to resort to analytic bounds on the error
to assess the performance of a parameter estimation technique. For the
mean-square error criterion, the Cram\'er-Rao bounds (CRBs) are the
most well known \cite{vantrees}. Although the CRBs are asymptotically
tight in the limit of infinitely many trials, it is well known that
the bounds can grossly underestimate the achievable error when the
likelihood function is highly non-Gaussian and the number of trials is
limited \cite{vantrees,bell}. For such situations, the Ziv-Zakai
bounds (ZZBs), which relate the mean-square error to the error
probability in a binary hypothesis testing problem, have been found to
be superior alternatives in many cases \cite{bell,zz}.  These bounds
are often much tighter in the highly non-Gaussian regime and can also
follow the CRBs closely for large numbers of trials \cite{bell}. In
physics, the ZZBs have also been applied to gravitational-wave
astronomy \cite{nicholson}.

The CRBs can be generalized for quantum parameter estimation, where
one estimates an unknown parameter such as phase shift, mirror
position, time, or magnetic field by measuring a quantum system such
as an optical beam, an atomic clock, or a spin ensemble
\cite{glm,helstrom,qcrb,twc}.  Given a quantum state to be measured,
the quantum CRBs (QCRBs) give error bounds that hold for any
measurement, but since they are always less tight to the error than
the corresponding classical CRBs \cite{braunstein}, the QCRBs share
all the shortcomings of their classical counterparts. This is an
outstanding problem in quantum metrology, as there have been many
claims based on the QCRBs or other similarly rudimentary arguments
about the parameter-estimation capabilities of certain exotic quantum
states \cite{ssw,dubious,rivas}, but such claims cannot be justified
if the bounds are not tight. Similar to the classical case, one
expects the QCRBs to be tight when many copies of the quantum object
are available \cite{glm}; the question is how many. For example,
Braunstein \textit{et al.}\ found numerically that the CRB for phase
estimation using the quantum state proposed by \cite{ssw} is tight
only when the number of copies exceeds a threshold \cite{blc}, while
Genoni \textit{et al.}\ found experimentally that the QCRB for a
phase-diffused coherent state is tight only after $\sim 100$ copies
have been measured \cite{genoni}.

In this Letter, I propose quantum Ziv-Zakai bounds (QZZBs) as
alternatives to the QCRBs for quantum parameter estimation.  The QZZBs
relate the mean-square error in a quantum parameter estimation problem
to the error probability in a quantum hypothesis testing problem, and
should be contrasted with previous studies that consider quantum
interferometry as a binary decision problem only \cite{paris}. To
demonstrate the versatility of the proposed bounds, I show that a
simple form of the bounds can produce both a ``Heisenberg'' error
limit (H limit \cite{hlimit_note}) that scales with the average energy
\cite{zwierz,hall,glm2} and another limit similar to the QCRB that
scales with the energy variance. I then illustrate these results by
applying the bound to a few examples of optical phase estimation. An
especially illuminating example is the state proposed by Rivas and
Luis, the QCRB of which can be arbitrarily low \cite{rivas}. I show
that a QZZB can be used to rule out any actual error scaling that is
better than the H-limit scaling for multiple copies of this
state. Beyond a certain number of copies, the QZZB starts to follow
the QCRB closely, thus revealing the regime where the QCRB must be
overly optimistic and indicating more precisely the asymptotic regime
where the QCRB is tight. Although the QZZBs are also lower error
bounds and not guaranteed to be tight either, the study here and the
usefulness of their classical counterparts suggest that they should be
similarly useful for quantum parameter estimation in general, whenever
one is suspicious about the tightness of the QCRBs.

Let $X$ be the unknown parameter, $Y$ be the observation, and $\tilde
X(Y)$ be an estimate of $X$ as a function of the observation
$Y$. Generalization to multiple parameters is possible \cite{bell} but
outside the scope of this Letter. The mean-square estimation error is
\begin{align}
\Sigma &\equiv \int dx dy P_{X,Y}(x,y)\Bk{\tilde X(y)-x}^2,
\end{align}
where $P_{X,Y}(x,y) = P_{Y|X}(y|x)P_X(x)$ is the joint probability
density of $X$ and $Y$, $P_{Y|X}(y|x)$ is the observation probability
density, also called the likelihood function when viewed as a function
of $x$, and $P_X(x)$ is the prior probability density. A classical ZZB
is given by \cite{bell}
\begin{align}
\Sigma &\ge \frac{1}{2}\int_0^\infty d\tau \tau
\mathcal V\intall dx \Bk{P_X(x)+P_X(x+\tau)}
\nonumber\\&\quad
\times \op{Pr}_e(x,x+\tau),
\label{zz1}
\end{align}
where $\op{Pr}_e(x,x+\tau)$ is the minimum error probability of the
binary hypothesis testing problem with hypotheses $\mathcal H_0: X =
x$ and $\mathcal H_1: X = x+\tau$, observation densities
$P_Y(y|\mathcal H_0) = P_{Y|X}(y|x)$, and $P_Y(y|\mathcal H_1) =
P_{Y|X}(y|x+\tau)$, and prior probabilities $P_0\equiv
\op{Pr}(\mathcal H_0) = P_X(x)/[P_X(x)+P_X(x+\tau)]$ and $P_1\equiv
\op{Pr}(\mathcal H_1) = 1-P_0$.  $\mathcal V$ denotes the optional
``valley-filling'' operation $\mathcal V f(\tau) \equiv \max_{\eta \ge
  0} f(\tau+\eta)$ \cite{bell}, which makes the bound tighter but more
difficult to calculate.  Another version of the ZZB is
\begin{align}
\Sigma &\ge \frac{1}{2}\int_0^\infty d\tau \tau
\mathcal V\intall dx 2\min \Bk{P_X(x), P_X(x+\tau)}
\nonumber\\&\quad
\times \op{Pr}_e^{el}(x,x+\tau),
\label{zz2}
\end{align}
where $\op{Pr}_e^{el}(x,x+\tau)$ is the minimum error probability of
the same hypothesis testing problem as before, except that the
hypotheses are now equally likely with $P_0 = P_1 = 1/2$.  If the
prior distribution $P_X(x)$ is a uniform window, the two bounds are
equivalent \cite{bell}.  For reference, Ref.~\cite{sup} includes
proofs of these bounds, following closely the ones in
Ref.~\cite{bell}.

To apply the bounds to the quantum parameter estimation problem, let
$\rho_X$ be the quantum state that depends on the unknown parameter
$X$ and $E(Y)$ be the positive operator-valued measure (POVM) that
models the measurement. The observation density becomes $P_{Y|X}(y|x)
= \trace\Bk{E(y)\rho_x}$.  The hypothesis testing problem then becomes
a state discrimination problem with the two possible states given by
$\rho_x$ and $\rho_{x+\tau}$. The error probability is bounded by a
lower limit first derived by Helstrom \cite{helstrom,fuchs}:
\begin{align}
\op{Pr}_e(x,x+\tau) &\ge 
\frac{1}{2}\bk{1-||P_0\rho_x-P_1\rho_{x+\tau}||_1},
\label{qzz}
\end{align}
where $||A||_1\equiv \trace\sqrt{A^\dagger A}$ is the trace
norm. Since all the quantities in the integral in Eq.~(\ref{zz1}) are
nonnegative, a lower quantum bound on the classical bound can be
obtained by replacing $\op{Pr}_e(x,x+\tau)$ in Eq.~(\ref{zz1}) with the
right-hand side of Eq.~(\ref{qzz}), resulting in a QZZB.  For
$\textrm{Pr}_{e}^{el}(x,x+\tau)$,
\begin{align}
\op{Pr}_e^{el}(x,x+\tau) &\ge 
\frac{1}{2}\bk{1-\frac{1}{2}||\rho_x-\rho_{x+\tau}||_1} 
\label{qzz_el}\\
&\ge
\frac{1}{2}\Bk{1-\sqrt{1-F(\rho_x,\rho_{x+\tau})}},
\label{qzz_fidel}
\end{align}
where $F$ is the quantum fidelity defined as $F(\rho_x,\rho_{x+\tau})
\equiv (\trace \sqrt{\sqrt{\rho_x}\rho_{x+\tau}\sqrt{\rho_x}})^2$.
The inequality in Eq.~(\ref{qzz_fidel}) is proved in Ref.~\cite{fuchs}
and becomes an equality when $\rho_X$ is pure.  For a product state
$\rho_X^{(1)}\otimes\rho_X^{(2)}\otimes\dots\rho_X^{(\nu)}$, $F =
\prod_{j=1}^{\nu}F(\rho_x^{(j)},\rho_{x+\tau}^{(j)})$, and
Eq.~(\ref{qzz_fidel}) is especially convenient.
Equations~(\ref{zz2}), (\ref{qzz_el}), and (\ref{qzz_fidel}) form
another QZZB, which is much more tractable and shall be used in the
remainder of the Letter. 

Similar to the Bayesian version of the QCRB \cite{qcrb,twc}, the QZZBs
allow one to compute lower quantum limits that hold for any
measurement and estimation method by considering only the quantum
state $\rho_X$ and the prior distribution $P_X(x)$.  There are,
however, at least three significant differences between the two
families of bounds: (1) The QZZBs are not expected to be saturable
exactly in general, unlike the QCRBs in special cases
\cite{bcm}, as the QZZBs are derived from the classical ZZBs,
which are also not saturable usually, and the Helstrom bounds, which
cannot be saturated for all $x$ and $\tau$ using one POVM. (2) While
the QCRBs depend only on the infinitesimal distance between $\rho_x$
and its neighborhood \cite{helstrom,braunstein}, the QZZBs depend on
the distance between $\rho_x$ and $\rho_{x+\tau}$ for all relevant
values of $x$ and $\tau$. (3) The QCRBs are ill-defined if $\rho_x$
and $P_X(x)$ are not differentiable with respect to $x$, whereas the
QZZBs have no such problem.

Assume now that $\rho_X$ is generated by the unitary evolution
\begin{align}
\rho_X &= \exp(-iHX)\rho\exp(iHX),
\end{align}
where $H$ is a Hamiltonian operator and $\rho$ is the initial
state. It can be shown that $F(\rho_x,\rho_{x+\tau}) \ge
\abs{\bra{\psi}\exp(-iH\tau)\ket{\psi}}^2$, where $\ket{\psi}$ is a
purification of $\rho$ with the same energy statistics
\cite{glm_speed}. Write $\ket{\psi}$ in the energy basis as
$\ket{\psi} = \sum_k C_k\ket{E_k}$ with $H\ket{E_k} = E_k\ket{E_k}$.
Then
\begin{align}
F(\rho_x,\rho_{x+\tau}) &\ge \bigg|\sum_k |C_k|^2\exp(-iE_k\tau)\bigg|^2
\nonumber\\
&= \sum_{k,l} |C_k|^2|C_{l}|^2 \cos[(E_k-E_{l})\tau]\equiv F(\tau),
\label{F0}
\end{align}
which is independent of $x$. Assume further that the prior
distribution is a uniform window with mean $\mu$ and width $W$
given by
\begin{align}
P_X(x) = \frac{1}{W}\rect\bk{\frac{x-\mu}{W}}.
\label{window}
\end{align}
With the optional $\mathcal V$ omitted, Eqs.~(\ref{zz2}),
(\ref{qzz_fidel}), (\ref{F0}), and (\ref{window}) give
\begin{align}
\Sigma &\ge \Sigma_{Z}\equiv \frac{1}{2}\int_0^W d\tau \tau\bk{1-\frac{\tau}{W}}
\Bk{1-\sqrt{1-F(\tau)}}.
\label{xiZ}
\end{align}
This inequality can be used to derive both an H limit and a QCRB-like
variance-dependent limit.

Applying the inequality $\cos\theta \ge 1-\lambda|\theta|$ to
Eq.~(\ref{F0}), where $\lambda\approx 0.7246$ is the implicit solution
of $\lambda = \sin \phi = (1-\cos \phi)/\phi$ for $0<\phi<\pi$, one
obtains $F(\tau) \ge \sum_{k,l} |C_k|^2|C_l|^2
\bk{1-\lambda|E_k-E_l|\tau}$.  Let $E_0$ be the minimum $E_k$. Then
$\Delta E_k\equiv E_k-E_0$ is nonnegative and $|E_k-E_l| = |\Delta
E_k-\Delta E_l|\le \Delta E_k+\Delta E_l$, which leads to
\begin{align}
F(\tau) &\ge  1-2\lambda H_+\tau,
& H_+ &\equiv \bra{\psi}H\ket{\psi}-E_0.
\label{Hp}
\end{align}
A tighter bound in terms of $H_+$ may be found using the formalism in
Ref.~\cite{glm_speed} but Eq.~(\ref{Hp}) suffices here. Since the
bound in Eq.~(\ref{Hp}) goes negative for $\tau > 1/(2\lambda H_+)$,
one can use the tighter bound $F(\tau)\ge 0$ there. Assuming a large
enough $H_+$ so that $W \ge 1/(2\lambda H_+)$, Eq.~(\ref{xiZ}) becomes
\begin{align}
\Sigma &\ge \Sigma_Z \ge \frac{1}{2}\int_0^{1/(2\lambda H_+)}
d\tau \tau\bk{1-\frac{\tau}{W}}
\bk{1-\sqrt{2\lambda H_+\tau}}
\nonumber\\
&= 
\frac{1}{80\lambda^2 H_+^2}-\frac{1}{336\lambda^3WH_+^3}
\nonumber\\
&\to \frac{1}{80\lambda^2 H_+^2}\textrm{ for }
W \gg\frac{1}{2\lambda H_+}.
\label{hlimit}
\end{align}
Equation~(\ref{hlimit}) is an H limit that scales with the average
energy relative to the ground state and does not depend on the prior
$W$ for large $H_+$. This result is subtly different from the one in
Ref.~\cite{glm2}, which does not average the mean-square error over a
prior distribution and uses a different method to prove the limit.
The limit derived in Ref.~\cite{hall}, on the other hand, does include
prior information and is tighter than Eq.~(\ref{hlimit}), but makes
the additional assumptions that $H$ has integer eigenvalues and $W =
2\pi$. The H limit derived here also does not contradict with
Ref.~\cite{boixo}, which assumes $H \propto n^k$, $k$ an integer, and
defines a different H limit in terms of $n$.

To derive another limit in terms of the energy variance, note that the
fidelity can also be bounded by \cite{glm_speed,fleming}
\begin{align}
F(\tau)
&\ge \cos^2(\Delta H\tau)
\textrm{ for }0\le \tau \le\frac{\pi}{2\Delta H},
\nonumber\\
\Delta H^2 &\equiv \bra{\psi}H^2\ket{\psi}
-\bra{\psi}H\ket{\psi}^2.
\end{align}
With $W \ge \pi/2\Delta H$, Eq.~(\ref{xiZ}) becomes
\begin{align}
\Sigma_Z &\ge \frac{1}{2}\int_0^{\pi/2\Delta H} d\tau 
\tau\bk{1-\frac{\tau}{W}}
\Bk{1-\sin(\Delta H\tau)}
\nonumber\\
&= \frac{\pi^2/16-1/2}{\Delta H^2}-\frac{1+\pi^3/48-\pi/2}{W\Delta H^3},
\nonumber\\
&\to \frac{\pi^2/16-1/2}{\Delta H^2}
 \textrm{ for }W \gg \frac{\pi}{2\Delta H},
\end{align}
which is less tight than the QCRB $\Sigma_C = 1/(4\Delta H^2)$ by a
constant factor but shows that the QZZB is also capable of predicting
the same scaling with the energy variance.

Consider now the problem of phase estimation using a harmonic
oscillator, assumed here to be an optical mode, with $H = n$, the
photon-number operator. For comparison, the Bayesian QCRB that
includes a prior Fisher information $\Pi\equiv
\int dx P_X(x)[\partial \ln P(x)/\partial x]^2$ is \cite{helstrom,qcrb,twc}
\begin{align}
\Sigma &\ge \Sigma_C \equiv\frac{1}{4\Delta N^2+\Pi},
\end{align}
where $\Delta N^2 \equiv \bra{\psi}n^2\ket{\psi}-N^2$ and $N\equiv
\bra{\psi}n\ket{\psi}$.  $\Pi$ is ill-defined for the prior given by
Eq.~(\ref{window}); I shall instead use a Gaussian prior distribution
with variance $W^2/12$ for the QCRB, so that $\Pi = 12/W^2$. For large
$\Delta N^2$, the prior information is irrelevant to the QCRB.  In
this regime, Ref.~\cite{sup} shows that the QZZB is less tight than
the QCRB by just a factor of 2 when the photon-number distribution
$|C_m|^2$ can be approximated as continuous and Gaussian, a case in
which the QCRB is known to be saturable \cite{bcm}. Thus the two
bounds can differ substantially only when $|C_m|^2$ is highly
non-Gaussian.

\begin{figure}[htbp]
\centerline{\includegraphics[width=0.49\textwidth]{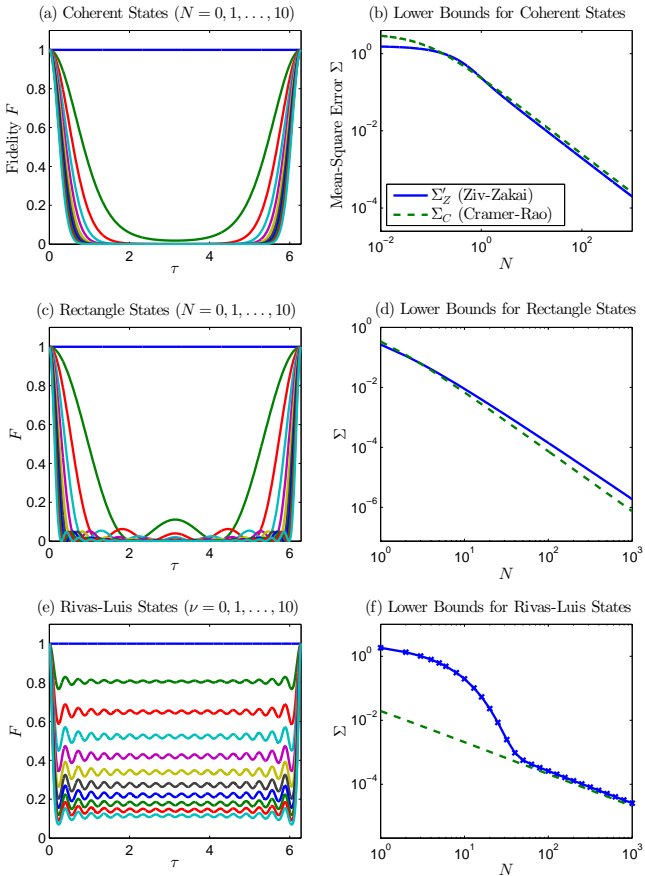}}
\caption{(Color online). Left column: fidelity of a pair of (a) coherent
  states, (c) rectangle states, and (e) products of Rivas-Luis states
  with $\epsilon = 0.1$ and $N_j = 1$ with phase difference
  $\tau$. Right column: mean-square-error lower bounds versus the
  average photon number on a log-log scale for (b) coherent states,
  (d) rectangle states, and (f) products of Rivas-Luis states with
  $\epsilon = 0.1$, $N_j = 1$, and $W = 2\pi$; straight lines
  connecting the numerically calculated points are guides for eyes.}
\label{zzs}
\end{figure}

Consider first a coherent state $\ket{\psi} = \exp(-N/2)
\sum_{m=0}^\infty (N^{m/2}/\sqrt{m!})\ket{m}$ with mean photon number
$N$.  $\Delta N^2 = N$, and the fidelity is $F(\tau,N) = \exp[2N(\cos
\tau-1)]$, as shown in Fig.~\ref{zzs}(a) for some different $N$'s.
For a product of coherent states, $\prod_{j = 1}^\nu F(\tau,N_{j}) =
F(\tau, \sum_{j=1}^\nu N_j)$ is identical to that for one coherent
state with the same total photon number on average.

For $W = 2\pi$, it can be shown \cite{sup} that Eq.~(\ref{xiZ}) gives
\begin{align}
\Sigma_{Z} &\ge \Sigma_{Z}' = \frac{\pi^{3/2}}{8\sqrt{N}}\exp(-4N)
\op{erfi}(2\sqrt{N}),
\label{zz_coh}
\end{align}
where $\op{erfi} z\equiv (2/\sqrt{\pi})\int_0^zdu \exp(u^2)$.  The
QZZB and the QCRB are plotted in Fig.~\ref{zzs}(b). In the limit of $N
\gg 1$, the right-hand side of Eq.~(\ref{zz_coh}) approaches
$\pi/(16N)$, which is slightly less than the QCRB given by $\Sigma_{C}
\to 1/(4N)$ but still obeys the expected $1/N$ ``shot-noise'' scaling.
%Since the QCRB is known to be tight for coherent states, it is not
%surprising to see that it does slightly better than the QZZB here.

Next, consider the state $\ket{\psi} =
(M+1)^{-1/2}\sum_{m=0}^M\ket{m}$, which has an equal superposition of
number states up to $\ket{M}$ and shall be called the rectangle state
here, with $N = M/2$ and $\Delta N^2 = M(M+2)/12$. The QCRB given by
\begin{align}
\Sigma_{C} = \frac{1}{4N(N+1)/3+\Pi}
\end{align}
follows the H-limit scaling $1/N^2$ for large $N$. The fidelity is
$F(\tau) = \sin^2[(2N+1)\tau/2]/[(2N+1)^2 \sin^2(\tau/2)]$.  Unlike
the coherent states, the fidelities for the rectangle states have
sidelobes, as shown in Fig.~\ref{zzs}(c).

The QZZB for $W =2\pi$ is \cite{sup}
\begin{align}
\Sigma_{Z} \ge \Sigma_Z' = \frac{\pi}{2(2N+1)^2}\sum_{k=0}^{2N}\frac{1}{2k+1}
\nonumber\\
\to \frac{\pi\ln (4N+1)}{4(2N+1)^2}
\textrm{ for large }N,
\end{align}
approaching a slower $\ln N/N^2$ scaling for large $N$. The additional
factor of $\ln N$ makes the QZZB diverge from the QCRB, as shown in
Fig.~\ref{zzs}(d). The $\ln N/N^2$ scaling was also observed
previously for the phase-squeezed state using other methods
\cite{collett}.

As the final example, consider the superposition of the vacuum with a
state $\ket{\zeta}$ that has a large photon-number variance, viz.,
$\ket{\psi}_j = c_0\ket{0} + c_1\ket{\zeta}$ with $|c_1|^2 \ll 1$, as
proposed by Rivas and Luis \cite{rivas}.  $\ket{\psi}_j$ can be
rewritten as $\ket{\psi}_j = \sqrt{1-\epsilon}\ket{0} +
\sqrt{\epsilon}\ket{\psi_s}$, where $\ket{\psi_s}$ is $\ket{\zeta}$
minus the vacuum component and renormalized. If $\ket{\psi_s}$ has a
mean photon number $N_s$ and photon-number variance given by $\gamma
N_s^2$ with $\gamma$ a constant, the mean and variance for
$\ket{\psi}_j$ are $N_j = \epsilon N_s$ and $\Delta N_j^2 =
[(1+\gamma)/\epsilon-1]N_j^2$.  $\Delta N_j^2$ can be made arbitrarily
larger than $N_j^2$ by reducing $\epsilon$, and the QCRB can be made
arbitrarily small. With $\nu$ copies and a total photon number $N
\equiv \nu N_j$,
\begin{align}
\Sigma_C &= \frac{1}{4[(1+\gamma)/(\nu\epsilon)-1/\nu]N^2 + \Pi}
\label{cr_rivas}
\end{align}
can decrease faster than the H-limit scaling $1/N^2$ if $\nu\epsilon$
also decreases with $N$ \cite{rivas}.

The fidelity tells a very different story. Defining the fidelity for
$\ket{\psi_s}$ as $F_s$, the fidelity for $\ket{\psi}_j$ is $F_j =
(1-\epsilon)^2 + 2\epsilon(1-\epsilon) \bra{\psi_s}\cos
n\tau\ket{\psi_s} +\epsilon^2 F_s \ge (1-\epsilon)^2 -
2\epsilon(1-\epsilon) = 1-4\epsilon+3\epsilon^2$.  Regardless of
$\ket{\psi_s}$, $F_j$ is bounded from below by a constant close to 1
if $\epsilon \ll 1$. For $\nu$ copies, $F = F_j^{\nu} \ge
(1-4\epsilon+3\epsilon^2)^\nu$, and a bound on the QZZB follows:
\begin{align}
\Sigma_Z &\ge \frac{W^2}{12}
\Bk{1-\sqrt{1-\bk{1-4\epsilon+3\epsilon^2}^\nu}}.
\label{zz_rivas}
\end{align}
This bound means that the actual error cannot deviate substantially
from the prior value $W^2/12$ until $\nu\epsilon \sim 1$, by which
point even if the error catches up with the QCRB it can no longer beat
the $1/N^2$ scaling. This result is unsurprising in light of the now
proven H limit.

To study the behavior of the Rivas-Luis state in more detail, let
$\ket{\psi}_j = \sqrt{1-\epsilon}\ket{0} +
\sqrt{\epsilon/M}\sum_{m=1}^M \ket{m}$.  Figure~\ref{zzs}(e) plots the
fidelities for some products of the Rivas-Luis states with $\epsilon =
0.1$ and $N = 1$, showing sharp features due to $\ket{\psi_s}$ near
$\tau = 0$ but quickly dropping off to the nonzero backgrounds due to
$\ket{0}$.  Figure~\ref{zzs}(f) plots the QZZB (calculated by
numerically integrating Eq.~(\ref{xiZ}) with $W = 2\pi$) and the QCRB
given by Eq.~(\ref{cr_rivas}) versus the total photon number $N$.  The
QZZB is much higher than the QCRB for small $N$ and comes down only
when $N \gtrsim 10$ and $\nu\epsilon \gtrsim 1$.  The QZZB then
reaches a threshold, beyond which it follows closely the QCRB.  This
threshold behavior is encountered frequently in classical parameter
estimation \cite{vantrees,bell} and also observed in a numerical study
of quantum phase estimation \cite{blc}.

In conclusion, the QZZBs are shown to be versatile error bounds that
can predict different types of quantum limits using one unified
formalism and can be much tighter than the popular QCRB for optical
phase estimation in certain cases. To model quantum sensors more
realistically, the QZZBs may also be generalized for waveform estimation in
a way similar to the QCRB \cite{twc}, if an error bound for continuous
quantum hypothesis testing \cite{tsang_hypo} can be found.

Discussions with L.~Maccone, V.~Giovannetti, M.J.W.~Hall, S.~Lloyd,
L.~Davidovich, B.M.~Escher, R.L.~de~Matos~Filho, M.G.A.~Paris,
\'A.~Rivas, A.~Luis, S.~Guha, A.~Tacla, C.M.~Caves, L.C.~Kwek,
A.~Ling, and J.P.~Dowling are gratefully acknowledged. This work is
supported by the Singapore National Research Foundation under NRF
Grant No. NRF-NRFF2011-07.

\appendix
\begin{widetext}
\section{Supplementary Material}
This Supplementary Material contains detailed derivations of some of
the results presented in the main text. Section~\ref{czzb} derives the
classical Ziv-Zakai bounds, Sec.~\ref{qzzb_gauss} calculates a quantum
Ziv-Zakai bound (QZZB) when the photon-number distribution can be
approximated as continuous and Gaussian, Sec.~\ref{qzzb_coh}
calculates the bound for coherent states, and Sec.~\ref{qzzb_rect}
calculates the bound for rectangle states.

\subsection{\label{czzb}Derivation of the classical Ziv-Zakai bounds}
Let
\begin{align}
\sigma &\equiv |\tilde X(Y)-X|
\end{align}
be a nonnegative random variable. The mean-square error becomes
\begin{align}
\Sigma &= \int_0^\infty ds P_\sigma(s)s^2,
\end{align}
where $P_\sigma(s)$ is the probability density of $\sigma$. With
\begin{align}
P_\sigma(s) &= -\diff{}{s}\op{Pr}\bk{\sigma \ge s}
\end{align}
and integration by parts, $\Sigma$ becomes
\begin{align}
\Sigma &= 2\int_0^\infty ds s\op{Pr}\bk{\sigma \ge s}
= \frac{1}{2}\int_0^\infty d\tau \tau\op{Pr}\bk{|\tilde X-X|\ge \frac{\tau}{2}}.
\end{align}
Since both $\tau$ and $\op{Pr}(|\tilde X-X|\ge \tau/2)$ are
nonnegative, one can find a lower bound on $\Sigma$ by bounding
$\op{Pr}(|\tilde X-X|\ge \tau/2)$.  Rewrite $\op{Pr}(|\tilde X-X|\ge
\tau/2)$ as follows:
\begin{align}
\op{Pr}\bk{|\tilde X-X|\ge \frac{\tau}{2}}
&= \op{Pr}\bk{\tilde X-X> \frac{\tau}{2}}
+\op{Pr}\bk{\tilde X-X \le -\frac{\tau}{2}}
\\
&= \intall dx_0 P_X(x_0)\op{Pr}\bk{\tilde X-X> \frac{\tau}{2}\bigg|X = x_0}
\nonumber\\&\quad
+
\intall dx_1 P_X(x_1)\op{Pr}\bk{\tilde X-X\le -\frac{\tau}{2}\bigg|X = x_1},
\end{align}
and let
\begin{align}
x_0 &= x, & x_1 &= x+\tau.
\end{align}
This yields
\begin{align}
\op{Pr}\bk{|\tilde X-X|\ge \frac{\tau}{2}}
&= \intall dx 
\bigg[P_X(x)\op{Pr}\bk{\tilde X> x+\frac{\tau}{2}\bigg|X = x}
\nonumber\\&\quad
+P_X(x+\tau)\op{Pr}\bk{\tilde X\le x+\frac{\tau}{2}\bigg|X = x+\tau}\bigg]
\label{eq1}
\\
&= \intall dx 
[P_X(x)+P_X(x+\tau)]\nonumber\\&\quad\times
\BK{P_0\op{Pr}\bk{\tilde X> x+\frac{\tau}{2}\bigg|X = x}
+
P_1\op{Pr}\bk{\tilde X\le x+\frac{\tau}{2}\bigg|X = x+\tau}},
\label{eq2}
\end{align}
where
\begin{align}
P_0 &\equiv \frac{P_X(x)}{P_X(x)+P_X(x+\tau)},
&
P_1 &\equiv \frac{P_X(x+\tau)}{P_X(x)+P_X(x+\tau)} = 1-P_0.
\end{align}
Now consider a binary hypothesis testing problem with hypotheses
\begin{align}
\mathcal H_0: X &= x, & \mathcal H_1: X &= x+\tau, 
\end{align}
prior probabilities
\begin{align}
\op{Pr}(\mathcal H_0) &= P_0,
&\op{Pr}(\mathcal H_1) &= P_1,
\end{align}
observation probability densities
\begin{align}
P_Y(y|\mathcal H_0) &= P_{Y|X}(y|x),
&
P_Y(y|\mathcal H_1) &= P_{Y|X}(y|x+\tau),
\end{align}
and the following suboptimal decision rule:
\begin{align}
\op{choose }\mathcal H_0 \op{ if }\tilde X &\le x + \frac{\tau}{2},
&
\op{choose }\mathcal H_1 \op{ if }\tilde X &> x + \frac{\tau}{2}.
\end{align}
The error probability of this hypothesis testing problem is then
precisely given by the expression in the curly brackets in
Eq.~(\ref{eq2}). This expression is lower-bounded by the minimum error
probability of the hypothesis testing problem, denoted by
$\op{Pr}_e(x,x+\tau)$, which does not depend on $\tilde X$, and one
obtains
\begin{align}
\op{Pr}\bk{|\tilde X-X|\ge \frac{\tau}{2}}
&\ge \intall dx [P_X(x)+P_X(x+\tau)] \op{Pr}_e(x,x+\tau).
\end{align}
The left-hand side is a monotonically decreasing function of $\tau$,
so a tighter bound can be obtained if we fill the valleys of the
right-hand side as a function of $\tau$.  Denoting this valley-filling
operation as $\mathcal V$:
\begin{align}
\mathcal V f(\tau) &\equiv \max_{\eta \ge 0} f(\tau+\eta),
\end{align}
one gets
\begin{align}
\op{Pr}\bk{|\tilde X-X|\ge \frac{\tau}{2}}
&\ge \mathcal V\intall dx 
[P_X(x)+P_X(x+\tau)]\op{Pr}_e(x,x+\tau),
\\
\Sigma &\ge \frac{1}{2}\int_0^\infty d\tau \tau
\mathcal V\intall dx 
[P_X(x)+P_X(x+\tau)]\op{Pr}_e(x,x+\tau).
\end{align}
This is a Ziv-Zakai bound. Another version that relates the
mean-square error to an equally-likely-hypothesis-testing problem can
be obtained from Eq.~(\ref{eq1}):
\begin{align}
\op{Pr}\bk{|\tilde X-X|\ge \frac{\tau}{2}}
&\ge \intall dx 2\min[P_X(x),P_X(x+\tau)]
\nonumber\\&\quad\times
\BK{\frac{1}{2}\op{Pr}\bk{\tilde X> x+\frac{\tau}{2}\bigg|X = x}
+\frac{1}{2}\op{Pr}\bk{\tilde X\le x+\frac{\tau}{2}\bigg|X = x+\tau}}.
\end{align}
The expression in the curly brackets is now the error probability of
the same hypothesis testing problem as before, except that the prior
probabilities are
\begin{align}
\op{Pr}(\mathcal H_0)=
\op{Pr}(\mathcal H_1) = \frac{1}{2}.
\end{align}
Another Ziv-Zakai bound follows:
\begin{align}
\Sigma &\ge \frac{1}{2}\int_0^\infty d\tau \tau\mathcal V\intall dx 2\min[P_X(x),P_X(x+\tau)]
\op{Pr}_{e}^{el}(x,x+\tau),
\end{align}
where $\op{Pr}_e^{el}(x,x+\tau)$ now denotes the minimum error
probability with equally likely hypotheses. Note that these bounds
make no assumption about the estimate.

% \section{A tighter ``Heisenberg'' limit}
% In general, the fidelity can be bounded by the inverse of an $\alpha$
% function defined in Ref.~\cite{glm_speed}:
% \begin{align}
% F(\tau) &\ge \alpha^{-1}\bk{\frac{2H_+\tau}{\pi}},
% \end{align}
% where the $\alpha$ has a tight lower bound given by
% \begin{align}
% \alpha(u) &\ge \min_\theta\max_q 
% \frac{2}{\pi b}\Bk{1-\sqrt{u}(\cos\theta-q\sin\theta)},
% &
% 0&\le u\le 1,
% \\
% \alpha(u) &= 0, & u &> 1,
% \end{align}
% and $b$ is defined implicitly in terms of $q$:
% \begin{align}
% b &= \frac{\phi+\sqrt{\phi^2(1+q^2)+q^2}}{1+\phi^2},
% &
% \sin \phi &= \frac{b(1-q\phi)+q}{1+q^2},
% &
% \pi - \tan^{-1}\frac{1}{q} \le \phi \le \pi + \tan^{-1}q.
% \end{align}
% The lower bound on $\alpha$ also provides a lower bound on
% $\alpha^{-1}$.  For $W \ge \pi/2H_+$, Eq.~(\ref{xiZ}) becomes
% \begin{align}
% \Sigma_Z &= \frac{1}{2}\int_0^{\pi/2H_+} d\tau \tau\bk{1-\frac{\tau}{W}}
% \Bk{1-\sqrt{1-\alpha^{-1}\bk{\frac{2H_+\tau}{\pi}}}}
% \\
% &= \frac{\pi^2}{8H_+^2}\int_0^{1} d\eta \eta\bk{1-\frac{\pi \eta}{2WH_+}}
% \Bk{1-\sqrt{1-\alpha^{-1}(\eta)}}.
% \end{align}
% According to Ref.~\cite{glm_speed},
% \begin{align}
% \alpha^{-1}(\eta) \approx \cos^2\frac{\pi\sqrt{\eta}}{2}
% \textrm{ for }
% 0\le \eta \le 1.
% \end{align}
% Then
% \begin{align}
% \Sigma_Z &\approx \frac{\pi^2}{8H_+^2}
% \int_0^{1} d\eta \eta\bk{1-\frac{\pi \eta}{2WH_+}}
% \bk{1-\sin\frac{\pi\sqrt{\eta}}{2}}
% \\
% &= \frac{\kappa^2}{H_+^2}-\frac{\chi}{WH_+^3},
% \quad
% \kappa \approx 0.2204,\quad
% \chi \approx 0.02770.
% \end{align}
\subsection{\label{qzzb_gauss}Quantum Ziv-Zakai bound for
approximately Gaussian photon-number distributions}
Consider the QZZB for optical phase estimation with
a uniform prior window:
\begin{align}
\Sigma_Z &\equiv \frac{1}{2}\int_0^W d\tau \tau\bk{1-\frac{\tau}{W}}
\Bk{1-\sqrt{1-F(\tau)}},
\\
F(\tau) &\equiv \Big|\sum_n |C_n|^2\exp(in\tau)\Big|^{2},
\label{fourier}
\end{align}
where $|C_n|^2$ is the photon-number distribution. If $|C_n|^2$ can be
approximated as a Gaussian distribution and the sum in
Eq.~(\ref{fourier}) as a continuous Fourier transform,
\begin{align}
|C_n|^2 &\approx \frac{1}{\sqrt{2\pi}\Delta N_j}
\exp\Bk{-\frac{(n-N_j)^2}{2\Delta N_j^2}},
\\
F (\tau) &\approx \Big|\int dn  |C_n|^2\exp(in\tau)\Big|^{2}
= \exp\bk{-\Delta N_j^2\tau^2}.
\end{align}
Since $1-\sqrt{1-F} \ge F/2$,
\begin{align}
\Sigma_Z &\gtrsim \frac{1}{4}\int_0^W d\tau \tau\bk{1-\frac{\tau}{W}}
F(\tau)
= \frac{1}{8\Delta N_j^2}
\textrm{ for } W\to\infty,
\end{align}
which is lower than the quantum Cram\'er-Rao bound (QCRB) by a constant
factor of 2. This shows that the two bounds can differ significantly
only when $|C_n|^2$ cannot be well approximated by a Gaussian.

For multiple copies, the fidelity can be written as
\begin{align}
F(\tau) &= \abs{\sum_{n_1,\dots,n_\nu} 
|C_{n_1}|^2|C_{n_2}|^2\dots |C_{n_\nu}|^2\exp\bigg(i\sum_j n_j \tau\bigg)}^{2},
\end{align}
which is the squared magnitude of the Fourier transform with respect
to the total photon number $\sum_j n_j$. By virtue of the central
limit theorem, the total-photon-number statistics will become
approximately Gaussian with variance $\nu\Delta N_j^2$ in the limit of
large $\nu$. This means that, regardless of the form of $|C_n|^2$, the
QZZB and the QCRB will become comparable in the limit of large $\nu$.

\subsection{\label{qzzb_coh}Quantum Ziv-Zakai bound for coherent states}
With the inequalities
\begin{align}
\tau\bk{1-\frac{\tau}{W}} &\ge \frac{W}{4}\sin\frac{\pi \tau}{W},
&
0&\le \tau\le W,
\\
1-\sqrt{1-F} &\ge \frac{1}{2}F,
&
0 &\le F \le 1,
\end{align}
the QZZB becomes
\begin{align}
\Sigma_Z &\equiv \frac{1}{2}\int_0^W d\tau \tau\bk{1-\frac{\tau}{W}}
\Bk{1-\sqrt{1-F(\tau)}}
\\
&\ge \frac{W}{16}\int_0^W d\tau \sin\frac{\pi \tau}{W}F(\tau).
\label{lowerZ}
\end{align}
For a coherent state and $W = 2\pi$,
\begin{align}
\Sigma_Z &\ge \frac{\pi}{8}\exp(-2N)
\int_0^{2\pi} d\tau \sin\frac{\tau}{2}\exp\bk{2N\cos \tau}.
\end{align}
Changing the integration variable to $u \equiv \cos(\tau/2)$ and using
the identity $\cos \tau = 2\cos^2(\tau/2)-1$, one obtains
\begin{align}
\Sigma_Z &\ge 
\frac{\pi}{2}\exp(-4N)
\int_{-1}^{1} du \exp(4N u^2),
\end{align}
which leads to Eq.~(16) in the text.

\subsection{\label{qzzb_rect}Quantum Ziv-Zakai bound for rectangle states}
Using Eq.~(\ref{lowerZ}) with $W = 2\pi$, one obtains
\begin{align}
\Sigma_Z &\ge \frac{\pi}{8(M+1)^2} I_{M},
\\
I_M &\equiv \int_0^{2\pi} d\tau \frac{\sin^2(M+1)\tau/2}{\sin \tau/2}.
\end{align}
It can be shown using trigonometric identities that
\begin{align}
\sin^2\frac{(M+1)\tau}{2}
&= \sin^2\frac{(M-1)\tau}{2} +
4\sin\frac{\tau}{2}\cos\frac{M\tau}{2}\sin\frac{(M-1)\tau}{2}
+4\sin^2\frac{\tau}{2}\cos^2\frac{M\tau}{2}.
\end{align}
The integral $I_M$ then satisfies the recursive relation
\begin{align}
I_M &= I_{M-2} + \frac{16M}{(2M-1)(2M+1)}
\\
&= I_{M-2} + 4\Bk{\frac{1}{2(M-1)+1}+\frac{1}{2M+1}},
\end{align}
with $I_{-1} = 0$ and $I_{0} = 4$. Hence
\begin{align}
I_M &= 4\sum_{k=0}^M\frac{1}{2k+1}.
\end{align}
For large $M$, the discrete sum can be approximated by an integral:
\begin{align}
I_M &\approx 4\int_0^M dk\frac{1}{2k+1} = 2\ln(2M+1).
\end{align}
\end{widetext}

\begin{thebibliography}{}
\bibitem{vantrees}H.~L.~Van Trees,
\textit{Detection, Estimation, and Modulation Theory, Part I}
(Wiley, New York, 2001).

\bibitem{bell}H.~L.~Van Trees and K.~L.~Bell (Eds.),
\textit{Bayesian Bounds for Parameter Estimation and Nonlinear Filtering/Tracking}
(Wiley-IEEE, Piscataway, 2007), and references therein.

\bibitem{zz}J.~Ziv and M.~Zakai,
IEEE Trans.\ Inform.\ Theor.\ \textbf{IT-15}, 386 (1969);
L.~P.~Seidman,
Proc.\ IEEE \textbf{58}, 644 (1970);
D.~Chazan, M.~Zakai, and J.~Ziv,
IEEE Trans.\ Inform.\ Theor.\ \textbf{IT-21}, 90 (1975);
S.~Bellini and G.~Tartara,
IEEE Trans.\ Commun.\ \textbf{COM-22}, 340 (1974);
E.~Weinstein,
IEEE Trans.\ Inform.\ Theor.\ \textbf{34}, 342 (1988).

\bibitem{nicholson}
D.~Nicholson and A.~Vecchio,
\prd \textbf{57}, 4588 (1998);
K.~J.~Lee \textit{et al.},
Mon.\ Not.\ R.\ Astron.\ Soc.\ \textbf{414}, 3251 (2011).

\bibitem{glm}V.~Giovannetti, S.~Lloyd, and L.~Maccone, 
Science  \textbf{306}, 1330 (2004);
Nature Photon. \textbf{5}, 222 (2011).

\bibitem{helstrom}C.~W.~Helstrom, 
\textit{Quantum Detection and Estimation Theory}
(Academic Press, New York, 1976);
A. S. Holevo,
\textit{Probabilistic and Statistical Aspects of Quantum Theory}
(North-Holland, Amsterdam, 1982);
H.~M.~Wiseman and G.~J.~Milburn,
\textit{Quantum Measurement and Control}
(Cambridge University Press, Cambridge, 2010).

\bibitem{qcrb}H.~P.~Yuen and M.~Lax,
IEEE Trans.\ Inform.\ Theor.\ \textbf{IT-19}, 740 (1973).

\bibitem{twc}M.~Tsang, H.~M.~Wiseman, and C.~M.~Caves,
\prl \textbf{106}, 090401 (2011).

\bibitem{braunstein}
This Letter defines the tightness of a bound by
comparing the bound to the achievable error.
A QCRB is always less tight to the error than the classical CRB
for a particular measurement strategy; see 
S.~L.~Braunstein and C.~M.~Caves,
\prl \textbf{72}, 3439 (1994).
This has motivated many studies
[see, for example, M.~G.~Genoni, S.~Olivares, and M.~G.~A.~Paris,
\prl \textbf{106}, 153603 (2011)] 
that analyze the tightness of a QCRB \emph{relative} to a classical CRB, 
but the QCRB cannot be tight to the error if the classical CRB is not.

\bibitem{ssw}
J.~H.~Shapiro, S.~R.~Shepard, and N.~C.~Wong,
\prl \textbf{62}, 2377 (1989);
J.~H.~Shapiro and S.~R.~Shepard,
\pra \textbf{43}, 3795 (1991).

\bibitem{dubious}
P.~M.~Anisimov \textit{et al.},
\prl \textbf{104}, 103602 (2010);
Y.~R.~Zhang \textit{et al.},
e-print arXiv:1105.2990.

\bibitem{rivas}\'A.~Rivas and A.~Luis,
e-print arXiv:1105.6310.

\bibitem{blc}S.~L.~Braunstein, A.~S.~Lane, and C.~M.~Caves,
\prl \textbf{69}, 2153 (1992).

\bibitem{genoni}
M.~Genoni \textit{et al.}, e-print arXiv:1203.2956.

\bibitem{paris}M.~G.~A.~Paris,
Phys.~Lett.~A \textbf{225}, 23 (1997);
Z.~Y.~Ou, \prl \textbf{77}, 2352 (1996).

\bibitem{hlimit_note} It is unfortunate that this limit has come to be
  known as the Heisenberg limit in the literature, as it is
  fundamentally different from the Heisenberg uncertainty relation,
  which is a relation of variances.

\bibitem{zwierz}
B.~Yurke, S.~L.~McCall, J.~R.~Klauder,
\pra \textbf{33}, 4033 (1986);
B.~C.~Sanders and G.~J.~Milburn, 
\prl \textbf{75}, 2944 (1995);
Z.~Y.~Ou, 
\prl \textbf{77}, 2352 (1996);
\pra \textbf{55}, 2598 (1997);
J.~J.~Bollinger, W.~M.~Itano, D.~J.~Wineland, 
and D.~J.~Heinzen,
\pra \textbf{54}, R4649 (1996);
P.~Hyllus, L.~Pezz\'e, and A.~Smerzi,
\prl \textbf{105}, 120501 (2010);
M.~Zwierz, C.~A.~Pe\'rez-Delgado, and P.~Kok,
\prl \textbf{105}, 180402 (2010);
\textbf{107}, 059904(E) (2011).

\bibitem{glm2}V.~Giovannetti, S.~Lloyd, and L.~Maccone,
e-print arXiv:1109.5661.

\bibitem{hall}
M.~J.~W.~Hall, D.~W.~Berry, M.~Zwierz, and H.~M.~Wiseman,
\pra \textbf{85}, 041802(R) (2012);
M.~J.~W.~Hall,
\jmo \textbf{40}, 809 (1993).

\bibitem{sup}See Supplementary Material for detailed calculations.

\bibitem{fuchs}C.~A.~Fuchs and J.~van~de~Graaf,
IEEE Trans.\ Inform.\ Theor.\ \textbf{45}, 1216 (1999).

\bibitem{bcm}S.~L.~Braunstein, C.~M.~Caves, and
G.~J.~Milburn,
Ann.\ Phys.\ \textbf{247}, 135 (1996).

\bibitem{glm_speed}
V.~Giovannetti, S.~Lloyd, and L.~Maccone,
\pra \textbf{67}, 052109 (2003).

\bibitem{boixo}S.~Boixo, S.~T.~Flammia, C.~M.~Caves,
and J.~M.~Geremia,
\prl \textbf{98}, 090401 (2007).

\bibitem{fleming} G.~H.~Fleming,
Nuovo Cimento \textbf{16A}, 232 (1973).

\bibitem{collett}M.~J.~Collett,
Physica Scripta \textbf{T48}, 124 (1993);
H.~M.~Wiseman and R.~B.~Killip,
\pra \textbf{57}, 2169 (1998);
D.~W.~Berry and H.~M.~Wiseman,
\pra \textbf{63}, 013813 (2000);
M.~Tsang, J.~H.~Shapiro, and S.~Lloyd,
\pra \textbf{78}, 053820 (2008); \textbf{79}, 053843 (2009).

\bibitem{tsang_hypo}M.~Tsang,
\prl \textbf{108}, 170502 (2012).
%e-print arXiv:1110.5058.

\end{thebibliography}
\end{document}